\documentclass[%
aip,jap,
 amsmath,amssymb,floatfix,
 preprint,%
]{revtex4-1}
\usepackage[utf8]{inputenc}
\usepackage[T1]{fontenc}
\usepackage{array}
\usepackage{amsmath,amssymb,amsfonts}
\usepackage{algorithmicx}
\usepackage{mathtools}
\usepackage{algorithm}
\usepackage{algpseudocode}
\usepackage{graphicx}
\usepackage{multirow}
\usepackage{textcomp}
\usepackage{xcolor}
\usepackage{siunitx}
\usepackage{mathtools}
\DeclarePairedDelimiter\abs{\lvert}{\rvert}%

\def\BibTeX{{\rm B\kern-.05em{\sc i\kern-.025em b}\kern-.08em
    T\kern-.1667em\lower.7ex\hbox{E}\kern-.125emX}}
\begin{document}

\title{2-input 4-output Programmable Spin Wave Logic Gate
}

\author{Abdulqader Mahmoud}
\email{a.n.n.mahmoud@tudelft.nl}
\affiliation{Delft University of Technology, Department of Quantum and Computer Engineering, 2628 CD Delft, The Netherlands}

\author{Frederic Vanderveken}
\affiliation{KU Leuven, Department of Materials, SIEM, 3001 Leuven, Belgium}
\affiliation{Imec, 3001 Leuven, Belgium}

\author{Christoph Adelmann}
\affiliation{Imec, 3001 Leuven, Belgium}

\author{Florin Ciubotaru}
\affiliation{Imec, 3001 Leuven, Belgium}

\author{Said Hamdioui}
\affiliation{Delft University of Technology, Department of Quantum and Computer Engineering, 2628 CD Delft, The Netherlands}

\author{Sorin Cotofana}
\email{S.D.Cotofana@tudelft.nl}
\affiliation{Delft University of Technology, Department of Quantum and Computer Engineering, 2628 CD Delft, The Netherlands}

\begin{abstract}
To bring Spin Wave (SW) based computing paradigm into practice and develop ultra low power Magnonic circuits and computation platforms, one needs basic logic gates that operate and can be cascaded within the SW domain without requiring back and forth conversion between the SW and voltage domains. To achieve this, SW gates have to possess intrinsic fanout capabilities, be input-output data representation coherent, and reconfigurable. In this paper, we address the first and the last requirements and propose a novel $4$-output programmable SW logic. First, we introduce the gate structure and demonstrate that, by adjusting the gate output detection method, it can parallelly evaluate any $4$-element subset of the $2$-input Boolean function set \{AND, NAND, OR, NOR, XOR, and XNOR\} . Furthermore, we adjust the structure such that all its $4$ outputs produce SWs with the same energy and demonstrate that it can evaluate Boolean function sets while providing fanout capabilities ranging from $1$ to $4$. We validate our approach by instantiating and simulating different gate configurations such as $4$-output AND/OR, $4$-output XOR/XNOR, output energy balanced $4$-output AND/OR, and output energy balanced $4$-output XOR/XNOR by means of Object Oriented Micromagnetic Framework (OOMMF) simulations. Finally, we evaluate the performance of our proposal in terms of delay and energy consumption and compare it against existing state-of-the-art SW and $16$~nm CMOS counterparts.  The results indicate that for the same functionality, our approach provides $3\times$ and $16\times$ energy reduction, when compared with conventional SW and $16$~nm CMOS implementations, respectively. 
\end{abstract}

\maketitle

\section{Introduction}
Processing the enormous amount of row data, which resulted due to the past decades information technology revolution, requires efficient computation platforms and CMOS downscaling has been sufficient to keep improving processing performance to match this requirements \cite{ITRS}. However, due to the technological difficulties: (i) leakage wall \cite{cmosscaling2}, (ii) reliability wall \cite{cmosscaling1}, and (iii) cost wall  \cite{cmosscaling2}, CMOS downscaling becomes very difficult at the nanoscale, which  eventually will soon lead to the end of Moore's law. Therefore, new technologies have been explored to find an alternative for CMOS,  e.g. memristors \cite{memristor}, graphene devices \cite{Yande1,Yande2}, and spintronics \cite{survey2}. Magnonics, a subset of spintronics, exploits Spin Waves (SWs) interactions to perform logic operations and appears to be a promising technology because of its attractive features \cite{survey1,survey2,ITRS}: (i) low power consumption as it doesn't need charge movement, (ii) acceptable delay, (iii) scalability down to the $nm$ range. 

Different spin wave logic gates have already been demonstrated \cite{logic21,logic12,logic11,logic17,logic25,logic4,logic16,logic18,logic24,Magnon_transistor, logic1, logic13,logic14,logic20, Abdulqader,logic19,logic2,logic3,logic100,logic101}. The first experimental spin wave logic gate was designed based on the Mach-Zender interferometer \cite{logic21}. Subsequently, XNOR, NAND, and NOR gates were designed using the same approach \cite{logic12,logic11,logic17}. Also, NOT, OR, and AND gates were built using three terminal transmission line devices \cite{logic25}\cite{logic4}\cite{logic16}\cite{logic18}. Different than the previously mentioned work, a nano-channel re-configurable spin wave device was employed to design voltage-controlled XNOR and NAND gates \cite{logic24}. Further, an XOR gate was designed by embedding two magnon transistors in the Mach-Zehnder interferometer arms \cite{Magnon_transistor}. In contrast to the aforementioned schemes, which encode information in amplitude, other proposals make use spin wave phase to encode the information \cite{logic1}. Consequently, buffer, inverter, (N)AND, (N)OR, XOR and Majority gates were built by embedding information in the spin wave phase and using both amplitude and phase to detect the information at the output \cite{logic1}. Additionally, different spin wave Majority gate geometries were suggested to decrease the back propagation of the spin waves \cite{logic13,logic14,logic20}. Also, a cross structure was used to design (N)OR gates \cite{logic19}. Several experimental results for Majority gates designs were also achieved \cite{logic2,logic3,logic100,logic101}. Despite these magnonics technology stpdf forward, state-of-the-art gates provide only one output thus they cannot provide fanout capabilities,  which are crucial for the effective implementation of large practically relevant circuits. Note that, if the output of such a gate should be fed to multiple following gates inputs, it must be multiple times replicated, which results in substantial area and energy consumption overheads. 

The problem of fanout is solved in this paper and multi-input multi-output Programmable Logic Gate (PLG) structures are proposed. The outputs can be the same or different depending on the design of the structure. This work main contributions are:\\
\begin{itemize}
  \item Development and design of a $2$-input $4$-output PLG, which can evaluate any $4$-element subset of the $2$-input Boolean function set \{AND, NAND, OR, NOR, XOR, and XNOR\}. For example, one such PLG gate can parallelly evaluate the set of $2$-input logic functions \{OR, NOR, XOR, XNOR\} on the same input combination.  
   \item Development and design of an output energy balanced $2$-input $4$-output PLG. The balanced $4$-output structure generates output SWs with the same energy, which implies intrinsic fanout capability. Therefore, the same function, e.g., X(N)OR, can be captured at different outputs and an up to $4$ fanout can be achieved without requiring gate replication.
  \item Functional validation and performance evaluation. We simulate different gate configurations, i.e., $4$-output AND/OR, $4$-output XOR/XNOR, output energy balanced $4$-output AND/OR, and output energy balanced $4$-output XOR/XNOR by means of OOMMF simulations.  We compare our proposal with the state-of-the-art work functionally equivalent counterparts and demonstrate that our approach provides $3\times$ and $16\times$ energy reduction, when compared with conventional SW and CMOS implementations, respectively.
\end{itemize}

The rest of the paper is organized as follows. Section \ref{sec:Basics of spin-wave technology} explains SW fundamentals and the associated computing paradigm. Section \ref{sec:Proposed programmable logic gate system design} introduces the proposed $2$-input $4$-output PLG structures. Section \ref{sec:Simulation Setup and Experiments} presents the simulation platform and utilized parameters. Section \ref{sec:Results and discussion} describes and provides the simulation results, and assesses the proposed structures against  current state-of-the-art designs. Also, fanout achievement, balance spin wave strength, variability and thermal noise effects are discussed. Section \ref{sec:Conclusion} concludes the paper.

\section{Spin-Wave Based Technology Basics}
\label{sec:Basics of spin-wave technology}
This section provides SW physics fundamentals and presents the SW interaction based  computation paradigm.

\subsection{Spin-Wave Fundamentals}
\label{sec:spin-wave fundamentals}

In a magnetic material, the magnetization can be exploited for memory or computation purposes. For example, in a magnetic equilibrium state, the magnetization is static which can be utilized to design spintronic memory devices. When the magnetization is out of equilibrium, it is subjected to a dynamical motion due to the magnetic torque.  The mathematical description of this magnetization dynamics is given by the Landau-Lifshitz-Gilbert (LLG) equation \cite{LL_eq}\cite{G_eq}:

\begin{equation} \label{eq:1}
\frac{d\vec{M}}{dt} =-\abs{\gamma} \mu_0 \left (\vec{M} \times \vec{H}_{eff} \right ) + \frac{\alpha}{M_s} \left (\vec{M} \times \frac{d\vec{M}}{dt}\right ),
\end{equation}
where $\gamma$ is the gyromagnetic ratio, $\alpha$ the damping factor, $M$ the magnetization, $M_s$ the saturation magnetization, and $H_{eff}$ the effective field which contains the different magnetic interactions
\begin{equation} \label{eq:2}
H_{eff}=H_{ext}+H_{ex}+H_{demag}+H_{ani},
\end{equation}
where $H_{ext}$ is the external field, $H_{ex}$ the exchange field, $H_{demag}$ the demagnetizing field, and $H_{ani}$ the magneto-crystalline field.

\begin{figure}[t]
\centering
  \includegraphics[width=\linewidth]{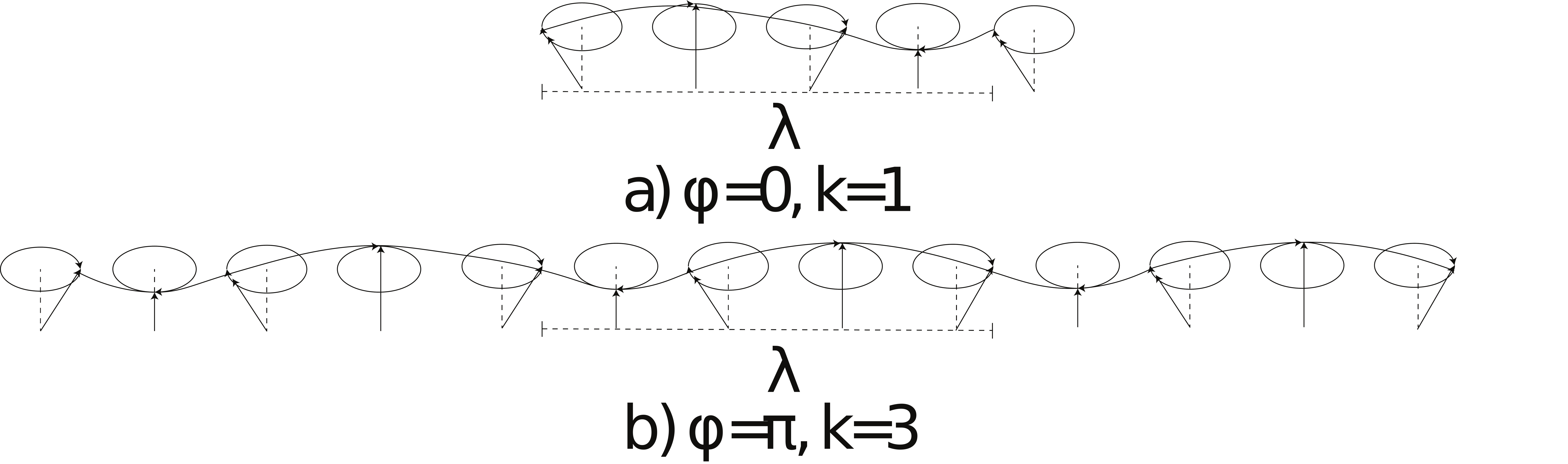}
  \caption{Spin Wave Parameters}
  \label{fig:SW_characterstics}
\end{figure}

For small magnetization perturbations Equation (\ref{eq:1}) can be linearised and has wave-like solutions. These weak wave-like solutions are called Spin Waves and can be seen as a collective excitation of the magnetization. Just like any other wave, a spin wave is completely described by its amplitude $A$, phase $\phi$, frequency $f$, wavelength $\lambda$, and wavenumber  $k=\frac{2\pi}{\lambda}$ as it can be observed in Figure \ref{fig:SW_characterstics}. The relation between the frequency $f$ and wavenumber $k$ is called the dispersion relation and is very important for the design of magnonic devices  \cite{dispersionrelation}. 

There are different SW types, each of which with its own properties. The static magnetization orientation with respect to the wave propagation direction determines which SW type gets excited  \cite{Magnetostatics_ref3}. In this work, Forward Volume Spin Waves (FVSW) are utilized, which corresponds to the case with static magnetization orientation out-of-plane. As a result, this type provides isotropic spin wave in plane propagation, which benefits circuits design that require different direction SW propagation \cite{Magnetostatics_ref3}. Note that this is not the case for the other SW types.

\begin{figure}[t]
\centering
  \includegraphics[width=\linewidth]{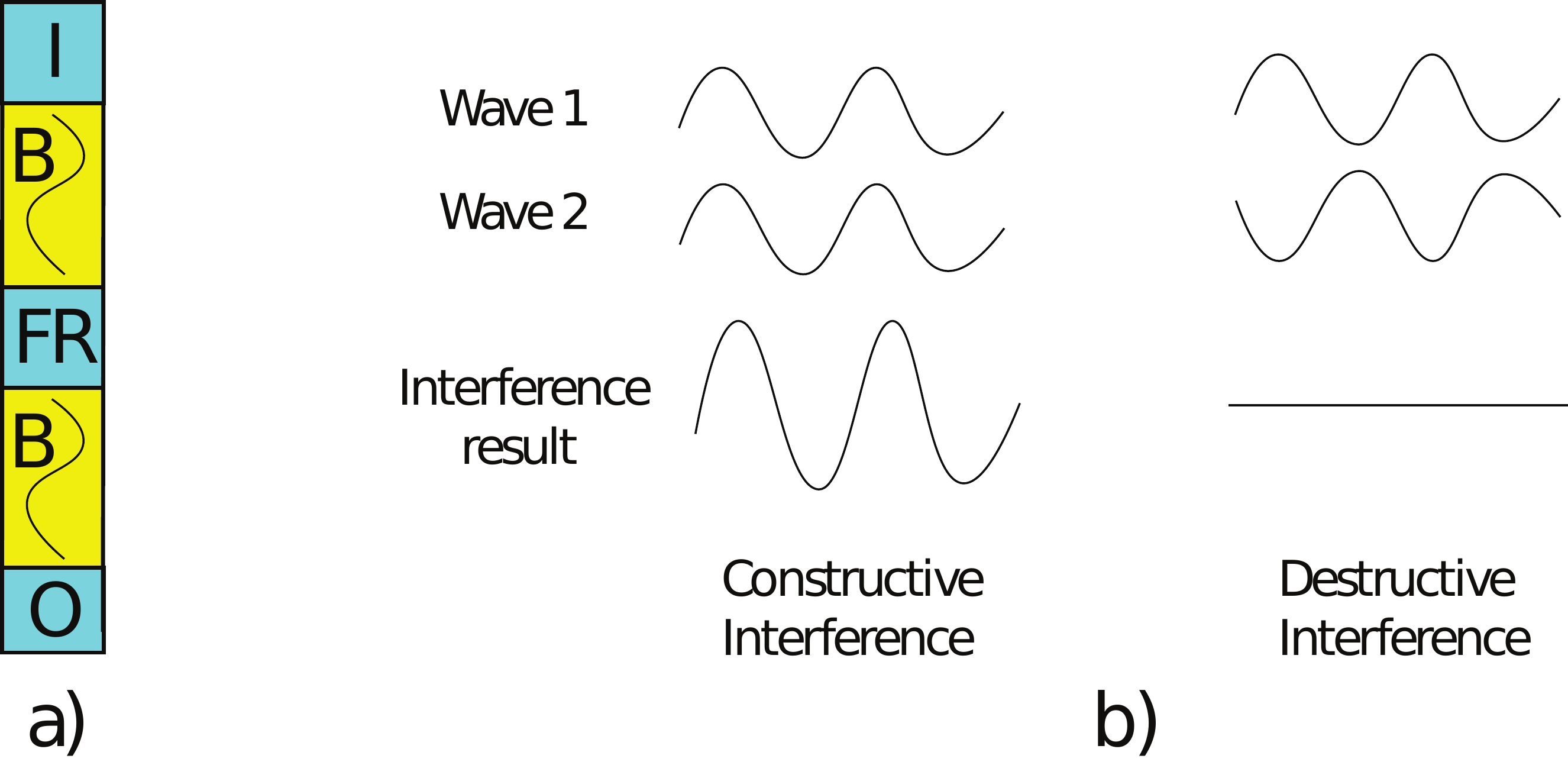}
  \caption{a) Spin Wave device b) Constructive and Destructive Interference}
  \label{fig:spin_wave_device}
\end{figure} 

\subsection{SW Computation Paradigm}
\label{sec:SW Computation Paradigm}

Figure \ref{fig:spin_wave_device}a presents a generic SW logic device that consists of four regions: excitation stage I, waveguide B, functional region FR, and detection stage O \cite{Magnonic_crystals_for_data_processing} . In the excitation stage SW are generated by means of microstrip antennas \cite{ref101,Magnonic_crystals_for_data_processing}, magnetoelectric cells \cite{ excitation1,excitation2} or spin orbit torque \cite{ref100,excitation4}). The waveguide is the medium for SW propagation and can be made of different magnetic materials, e.g., Permalloy, Yttrium iron garnet, CoFeB \cite{Magnonic_crystals_for_data_processing}. The selected waveguide material is an important choice as this fundamentally  influences the SW properties. In the functional region, SWs can be amplified, normalized or interfere with other SWs. In the detection stage, the spin wave is captured and converted to the electrical domain via microstrip antennas \cite{ref101,Magnonic_crystals_for_data_processing}, magnetoelectric cells \cite{ excitation1,excitation2} or spin orbit torque \cite{ref100,excitation4}.

During SW excitation, its amplitude and phase can be utilized to encode information. This can be done simultaneously at different spin wave frequencies \cite{parallel_data_processing1}, which potentially allows for parallel data processing. The interaction between SWs in the same waveguide is governed by the interference principle. To explain the interference principle, we make use of two SW interference as discussion vehicle. The interference result is constructive when they have the same phase $\Delta \phi=0$, whereas if they are out of phase $\Delta \phi=\pi$, the interference is destructive as depicted in Figure \ref{fig:spin_wave_device}b. Consequently, if more than two waves coexist in the waveguide, the majority principle governs the interference result. For example, if 3 SWs are present in a waveguide and at most one SW has phase $\pi$  while the others have phase $0$  the interference result will be a SW with $\phi=0$, whereas a SW with $\phi=\pi$ will be the result if two or all SWs have phase $\pi$.  We note that the implementation of such a $3$-input Majority gate in CMOS requires $18$ transistors \cite{logic1,logic9}, while it can be directly implementation in the SW domain by the interference of $3$ SWs. More complex interference cases exist if the propagated SWs have different $A$, $\lambda$, and $f$, which might be of interest for designing novel magnonic computing systems. However, in this paper, we only focus on the simplest case, i.e., excited SWs have the same $A$, $\lambda$, and $f$ and can take two discrete phases $\phi=0$ and $\phi=\pi$. Logic $0$ refers to a SW with $\phi=0$, and a Logic $1$ refers to a SW with $\phi=\pi$.

\section{$2$-input $4$-output Programmable Logic Gate}
\label{sec:Proposed programmable logic gate system design}
This section introduces the novel $2$-input  $4$-output programmable logic SW gate structures.

\subsection{Unbalanced $2$-input $4$-output Programmable Logic Gate}

\begin{figure}[t]
\centering
  \includegraphics[width=0.3\linewidth]{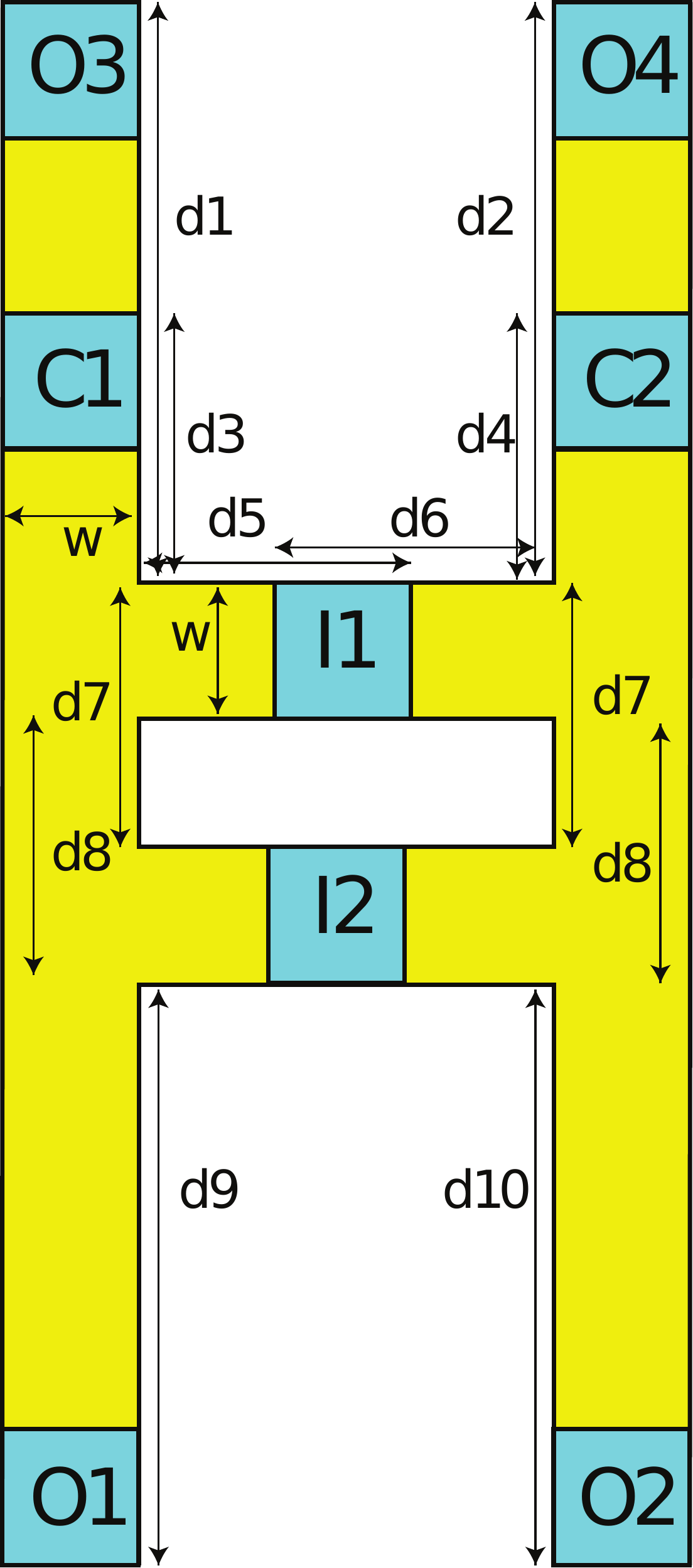}
  \caption{$2$-input $4$-output SW Programmable Logic Gate}
  \label{fig:FO4}
\end{figure}

Figure \ref{fig:FO4} presents the proposed $2$-input $4$-output PLG. The structure has a ladder shape with two data inputs $I_1$ and $I_2$, and two controls inputs $C_1$ and $C_2$. The outputs $O_1$, $O_2$, $O_3$, and $O_4$ correspond to the detection cells where the gate results are obtained. The excitation and detection stages can be voltage-encoded (or current-encoded) depending on the utilized excitation/detection method. As mentioned previously, there are multiple options for the SWs excitation and detection, e.g., microstrip antennas \cite{ref101,Magnonic_crystals_for_data_processing}, magnetoelectric cells \cite{ excitation1,excitation2,excitation2,excitation3}, spin orbit torque \cite{ref100,excitation4}.

In principle, the structure is generic and functions correctly if the input SWs have the same amplitude $A$, wavelength $\lambda$, and frequency $f$ regardless of their values, while the chosen $A, \lambda,$ and $f$ values determine its dimensions. %However, to simplify the calculation, SWs must be excited with the same amplitude, wavelength and frequency. 
To guarantee a proper behaviour, the structure dimensions must be precisely determined. For example, if SWs should interfere constructively when they have the same phase and destructively for opposite phases the dimensions must be $d_3=d_4=d_5=\ldots=d_8= n \times \lambda$ (where n=0, 1, 2, 3, \ldots). When the opposite behaviour is desired, SWs interfering constructively when they are out of phase and destructively when they are in phase, then the dimensions should be $d_3=d_4=d_5=\ldots=d_8= (n+\frac{1}{2}) \times \lambda$.

\begin{table}
\caption{$2$-input Programmable Gate Behaviour}
\label{table:9}
\centering
  \begin{tabular}{|c|c|c|c|c|c|c|}
    \hline
    Output detection method &  $C_1$ &  $C_2$ &  $O_1$ &  $O_2$ &  $O_3$ &  $O_4$ \tabularnewline
    \hline
    \multirow{4}{8em}{Output detection by SW phase} &  $0$ &  $0$ & (N)AND & (N)AND  & (N)AND  & (N)AND  \tabularnewline \cline{2-7}
    &  $0$ &  $1$ & (N)AND  & (N)OR & (N)AND  & (N)OR \tabularnewline \cline{2-7}
    &  $1$ &  $0$ & (N)OR & (N)AND  & (N)OR & (N)AND  \tabularnewline \cline{2-7}
     &  $1$ &  $1$ & (N)OR & (N)OR & (N)OR & (N)OR \tabularnewline 
    \hline
    \multirow{4}{8em}{Output detection by thresholding} &  $0$ &  $0$ &  X(N)OR & X(N)OR  &  - & - \tabularnewline \cline{2-7}
   &  $0$ &  $1$ & X(N)OR & X(N)OR & - & - \tabularnewline \cline{2-7}
    &  $1$ &  $1$ & X(N)OR & X(N)OR  & - & - \tabularnewline \cline{2-7}
    &  $1$ &  $1$ & X(N)OR & X(N)OR  & - & - \tabularnewline
    \hline
  \end{tabular}
\end{table}

Moreover, two ways of output detection exist: (i) Phase Detection (PD) and (ii) Threshold Detection (TD). Depending on a predefined phase, PD is performed as follows: if output SW phase is $\phi=0$ the output is logic $0$ and logic $1$ if $\phi=\pi$. For TD the Magnetization Spinning Angle (MSA) is measured and compared with a predefined threshold value such that if MSA  is larger than the threshold, the output is logic $1$, and logic $0$ otherwise. The MSA is calculated as 
 \begin{equation} \label{eq:3}
MSA = \arctan \left(\frac{\sqrt{(\overline{m_x})^2+(\overline{m_y})^2}}{M_s}\right),
\end{equation}
where $\overline{m_x}$ and $\overline{m_x}$ are the mean of the magnetization on $x$ and $y$ directions, respectively. 

Further, the position of the PLG outputs $O_1$, $O_2$, $O_3$, and $O_4$ must be also  accurately determined to obtain the desired results at the outputs. If the used detection method is phase detection, the result can be the logic gate output itself or its inverted version depending on the position. If the direct logic function is of interest, the distances must be $d_{1}=d_{2}=d_{9}=d_{10}= n \times \lambda$ whereas if the complement is of interest, the distances should be $d_{1}=d_{2}=d_{9}=d_{10}= (n+\frac{1}{2}) \times \lambda$. Whereas if the used detection method is threshold detection, the output should be as close as possible from the last interference point to have strong spin wave. In this case, if the complement is of interest, then the aformentioned condition can be flipped such that if MSA is less than the threshold, the output is logic $1$, and logic $0$ otherwise.
 
Table \ref{table:9} summarizes the logic gates behaviors of the structure in Figure \ref{fig:FO4} depending on the control signal values and the output detection methods. If PD is utilized, (N)AND and/or (N)OR gates are implemented. In contrast, XOR and/or XNOR gates are obtained if TD is utilized. Furthermore, both mechanisms can be utilized in the same time, i.e., some outputs can PD captured and others TD captured. Therefore, some outputs can be (N)AND or (N)OR gate and the others can be XOR or XNOR gate. However, the XOR/XNOR functionality cannot be obtained at $O_3$ and $O_4$ because they receive amplitude unbalanced SW due to the fact that $I_3$ and $I_4$ are closer to $O_3$ and $O_4$ than to $O_1$ and $O_2$. The unbalance SW amplitude also causes unbalance in the output energy as it clarified in Section \ref{sec:Results and discussion}. Therefore, to enable full gate flexibility a balance PLG design is needed as will be introduced in the following subsection. 
Depending on the desired functionality the PLG can simultaneously evaluate up to $4$ $2$-input different basic Boolean functions. Note that the structure in Figure \ref{fig:FO4} can be extended and can have multiple inputs. 

To illustrate the PLG operation principle, we consider a $2$-input AND/OR $C_1=0$ and $C_2=1$ with phase based output detection. All excited SWs have the same amplitude, frequency, and wavelength. If logic $0$ is applied on both inputs $I_1$ and $I_2$, then SWs with phase of $0$ are excited at $I_1$ and $I_2$. The SWs propagate in both sides of the excitation cells. Once the SW excited at $I_1$ arrives to the left arm, it  constructively interferes with the SW excited at $C_1$. Then, the result propagates and interferes constructively with the SW excited at $I_2$. The result of interference further propagates to be captured at $O_1$. The captured result is logic $0$ as the interferences resulted in a SW with phase of $0$ at the output. The same result is captured at $O_3$ as the spin wave propagates in both directions. In the case when logic $0$ is applied on $I_1$ and logic $1$  on $I_2$ the SW excited at $I_1$ constructively interferes with the SW excited at $C_1$. The resulted SW propagates to interfere destructively with the SW excited at $I_2$. The result obtained from the interference is a SW with phase of $0$, which is captured at $O_1$ and $O_3$ resulting in a logic $0$. If logic $1$ is applied on $I_1$ and logic $0$ is applied on $I_2$, then the SW excited at $I_1$ interferes destructively with the SW excited at $C_1$, which results in a very low SW energy (if not vanishing each other). Thus, the only SW in the device is the one excited from $I_2$, which propagates to the outputs $O_1$ and $O_3$ and captured as a logic $0$ because it has a $0$ phase. Finally, if logic $1$ is applied on both inputs $I_1$ and $I_2$, then it is similar to the previous case and the SW excited from $I_2$ is the only SW in the waveguide. Thus, the captured result at the outputs $O_1$ and $O_3$ is logic $1$ because the resulted SW has a $\pi$ phase. The same analysis can be followed to reach to the result in gates' right arm.

If TD is utilized to capture the results the performed function becomes XOR/XNOR. In this approach, the output SW phase is ignored and  MSA or amplitude is the desired information. Therefore, if output MSA is greater than the threshold a logic $1$ is generated and a logic $0$ otherwise. Following this approach an XOR gate is implemented. If XNOR is desired, then condition must be flipped, i.e., the output is logic $1$ when the MSA is less than the threshold and the output is logic $0$, otherwise.

\subsection{Balanced $2$-input  $4$-output Programmable Logic Gate}

\begin{figure}[t]
\centering
  \includegraphics[width=0.3\linewidth]{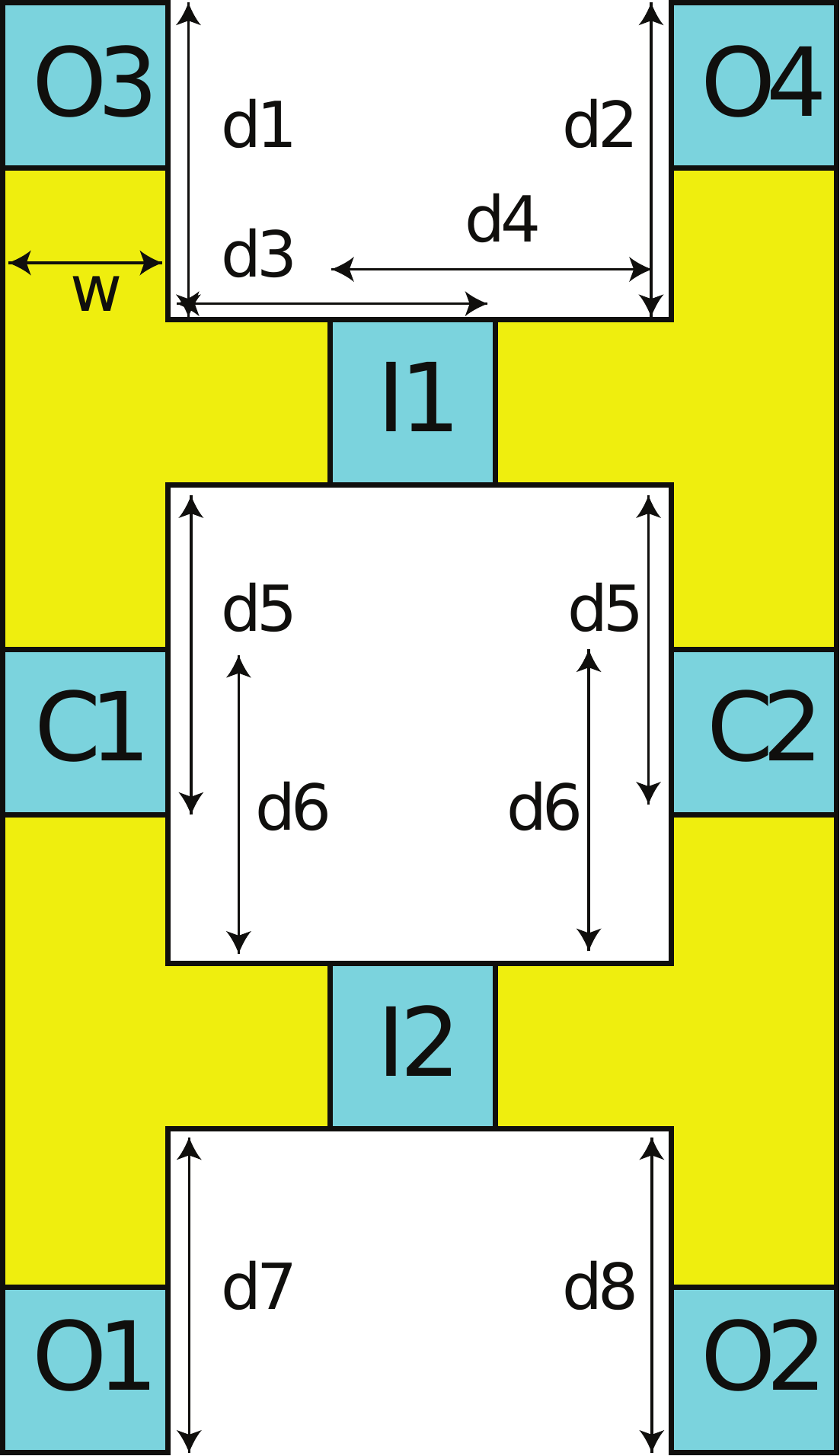}
  \caption{$2$-input $4$-output Output Energy Balanced SW Programable Logic Gate}
  \label{fig:OEBFO4}
\end{figure} 

\begin{table}
\caption{2-input Programmable Gate Behaviour}
\label{table:10}
\centering
  \begin{tabular}{|c|c|c|c|c|c|c|}
    \hline
    Output detection method &  $C_1$ &  $C_2$ &  $O_1$ &  $O_2$ &  $O_3$ &  $O_4$ \tabularnewline
    \hline
    \multirow{4}{8em}{Output detection by SW phase} &  $0$ &  $0$ & (N)AND & (N)AND  & (N)AND  & (N)AND  \tabularnewline \cline{2-7}
    &  $0$ &  $1$ & (N)AND  & (N)OR & (N)AND  & (N)OR \tabularnewline \cline{2-7}
    &  $1$ &  $0$ & (N)OR & (N)AND  & (N)OR & (N)AND  \tabularnewline \cline{2-7}
     &  $1$ &  $1$ & (N)OR & (N)OR & (N)OR & (N)OR \tabularnewline 
    \hline
    \multirow{4}{8em}{Output detection by thresholding} &  $0$ &  $0$ &  X(N)OR & X(N)OR  &  X(N)OR & X(N)OR \tabularnewline \cline{2-7}
   &  $0$ &  $1$ & X(N)OR & X(N)OR & X(N)OR & X(N)OR \tabularnewline \cline{2-7}
    &  $1$ &  $1$ & X(N)OR & X(N)OR  & X(N)OR & X(N)OR\tabularnewline \cline{2-7}
    &  $1$ &  $1$ & X(N)OR & X(N)OR  & X(N)OR & X(N)OR \tabularnewline
    \hline
  \end{tabular}
\end{table}

As previously mentioned due to the lack of symmetry the $4$ outputs are not fully equivalent in terms of computation capabilities. To circumvent this limitation we proposed a symmetric energy balanced $4$-input PLG depicted in Figure \ref{fig:OEBFO4}. To balance the output energies and be able to capture the result of all possible logic functions at all outputs, we relocate the control inputs in the middle of the vertical waveguide such that each gate input is located at the same distance from all the four gate outputs. Therefore, the waves propagate towards $O_1$, $O_2$, $O_3$, and $O_4$ on equal length paths, which means that the rich the outputs with the same (amplitude) energy.  The previously described design procedures is in place and all logic functions are feasible at each outputs as demonstrated by Table \ref{table:10}. An extra advantage of this structure is that when computing the same function it can provide a clean maximum fanout of $4$, or when computing $2$ functions each of them can be produced with a fanout of $2$.

\section{Simulation Setup}
\label{sec:Simulation Setup and Experiments}
In the following lines an overview of the simulation platform and used parameters are provided.

The Object Oriented MicroMagnetic Framework (OOMMF) is a micromagnetic simulator, which solves numerically the LLG equation \cite{OOMMF} . This software is used to validate the proposed structures.  $Fe_{60}Co_{20}B_{20}$ is utilized as waveguide material and its parameters are presented in Table \ref{table:2} \cite{parameters}. The width of the waveguide is $50$ nm and the thickness is $1$ nm. The static magnetization is out-of-plane by Perpendicular Magnetic Anisotropy (PMA) and no  external magnetic field is require as the PMA field is larger than the magnetic saturation \cite{parameters}. The spin wave wavelength of  $\lambda = 110$ nm is chosen to be larger than the width of the waveguide. Once, the wavelength is determined, the distances can be calculated and become $ d_3 = d_4 = d_5 = d_6= d_7= d_8 = 110$ nm for the structure in Figure \ref{fig:FO4} and $ d_3 = d_4 = d_5 = d_6 = 110$ nm for the structure in Figure \ref{fig:OEBFO4}. Also, as $\lambda =110$ nm, and $k=2\pi/\lambda=57$ rad/$\mu$m, the SW frequency becomes $f = 9$ GHz according to the dispersion relation.
\begin{table}[t]
\caption{Parameters}
\label{table:2}
\centering
  \begin{tabular}{|c|c|}
    \hline
    Parameters & Values \\
    \hline
    Magnetic saturation $M_s$ & $1.1$ $\times$ $10^6$ A/m \\
    \hline
    Perpendicular anisotropy constant $k_{ani}$ & $8.3177$ $\times$ $10^5$ J/$m^3$\\
    \hline
    damping constant $\alpha$ & $0.004$ \\
    \hline
    Waveguide thickness $t$ & $1$ nm \\
    \hline
    Exchange stiffness $A_{exch}$ & $18.5$ pJ/m \\
    \hline
  \end{tabular}
\end{table}

\section{Results and Discussion}
\label{sec:Results and discussion}
This section provides OOMMF simulation results and gate performance evaluation and comparison with equivalent state-of-the-art SW and CMOS counterparts. A discussion about fanout achievement, balance spin wave strength and variability and thermal noise effect issues are also included.

\subsection{Simulation Results}
\label{subsec:Results}

\subsubsection*{\textbf{$2$-input $4$-output AND/OR gates}}

\begin{figure}[t]
\centering
  \includegraphics[width=\linewidth]{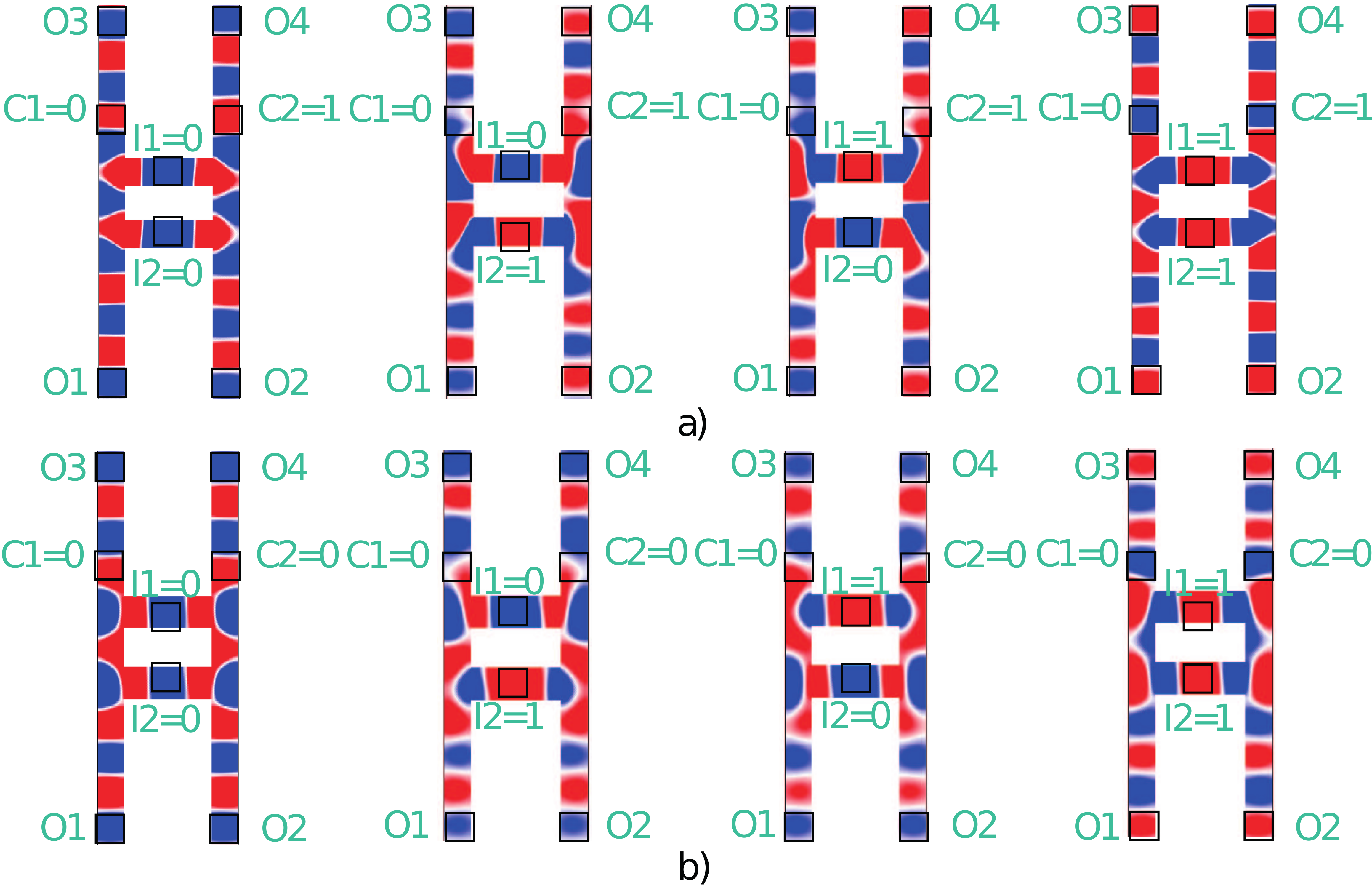}
  \caption{$2$-input $4$-output a) AND/OR Gate b) AND/AND Gate}
  \label{fig:results3}
\end{figure}

Figure \ref{fig:results3}a presents simulation results for the $2$-input $4$-output AND/OR gates for $4$ cases  $I_1I_2$ $=$ {$00$ $01$ $10$ $11$}. The outputs $O_1$, $O_2$, $O_3$, and $O_4$ are placed at  $d_1$$=$$d_2$$=$$d_9$$=$$d_{10}$$=$$220$nm (n=2). Also simulation results for $2$-input AND/AND gates are presented in Figures \ref{fig:results3}b. As it is clear in Figure \ref{fig:results3}a that the left arm provides the AND gate functionality at outputs $O_1$ and $O_3$, whereas the right arm provides the OR gate functionality at outputs $O_2$ and $O_4$. Taking $O_1$ and $O_3$ as an example, if inputs $I_1I_2$$=$$00$,  $I_1I_2$$=$$01$,  $I_1I_2$$=$$10$, then the output $O_1$$=$$0$ and $O_3$$=$$0$. In contrast, $O_1$$=$$1$ and $O_3$$=$$1$ for the input combination $I_1I_2$$=$$11$. The OR gate functionality is obtained from $O_2$ and $O_4$. In addition, NAND and NOR gates can be captured by changing the reading positions to be at $3\lambda/2$, i.e., $d_1$$=$$d_2$$=$$d_9$$=$$d_{10}$=$165$nm (n=1). Likewise, Figure \ref{fig:results3}b can be analyzed. Also, $2$-input (N)OR/(N)OR gates can be obtained in the same manner but with $C_1$$=$$C_2$$=$$1$. Therefore, the structure can provide AND, NAND, OR, and NOR gate functionalities while each gate column being able to provide  AND (OR) in its direct and inverted format or in the same format with a fanout of $2$.

\subsubsection*{\textbf{$2$-input $4$-output XOR/XNOR gates}}

Table \ref{table:5} presents normalized MSAs at the outputs $O_1$, $O_2$, $O_3$, and $O_4$ for $C_1$$=$$C_2$$=$$0$ and $C_1$$=$$C_2$$=$$1$ and for different inputs combination $I_1I_2$ $=$ {$00$ $01$ $10$ $11$} for the structure in Figure \ref{fig:FO4}. Note that the results for the cases $C_1 C_2$ $=$ {$01$ $10$} are exhibiting the same behaviour.

Table \ref{table:5} indicate that the outputs $O_1$ and $O_2$ can provide XOR or XNOR logic gates if an appropriate threshold is set to detect logic $0$ and logic $1$. On the other hand, $O_3$ and $O_4$ cannot provide these logic gates. As it can be observed from the table, the XOR gate can be implemented at $O_1$ and $O_2$ by averaging the $O_1$ and $O_2$ normalized MSAs for input combinations 10 and 11, which is $0.35$. The XOR gate can be obtained by setting the condition that the normalized MSA is greater than $0.35$ for logic $0$ and logic $1$, otherwise. By reversing the condition, the XNOR gate is obtained at $O_1$ and $O_2$. As it is clear from the Table, the four outputs don't have the same MSA and cannot provide XOR and XNOR functionalities (only $O_1$ and $O_2$ can). Thus, to balance the output energies and to enable XOR and XNOR in all four outputs, we place the control inputs as depicted in Figure \ref{fig:OEBFO4}.

\begin{table}[t]
\caption{$2$-input $4$-output Gate Normalized Outputs MSAs}
\label{table:5}
\centering
  \begin{tabular}{|c|c|c|c|c|c|c|}
    \hline
   \multicolumn{3}{|c|}{Cases} & $O_1/I_1$  & $O_2/I_1$  & $O_3/I_1$  & $O_4/I_1$ \tabularnewline \hline
    $C_1=C_2$ & $I_2$& $I_1$& & & & \tabularnewline
    \hline
    \centering $0$ & $0$ & $0$ & $0.9$ & $0.9$ & $1$ & $1$\tabularnewline
    \hline
    \centering $0$ & $0$ & $1$ & $0.25$ & $0.25$ & $0.45$ & $0.43$\tabularnewline
    \hline
    \centering $0$ & $1$ & $0$ & $0.32$ & $0.32$ & $0.26$ & $0.27$ \tabularnewline
    \hline
    \centering $0$ & $1$ & $1$ & $0.38$ & $0.39$ & $0.33$ & $0.33$ \tabularnewline
    \hline
    \centering $1$ & $0$ & $0$ & $0.38$ & $0.39$ & $0.33$ & $0.33$ \tabularnewline
    \hline
    \centering $1$ & $0$ & $1$ & $0.32$ & $0.32$ & $0.26$ & $0.27$  \tabularnewline
    \hline
    \centering $1$ & $1$ & $0$ & $0.25$ & $0.25$ & $0.45$ & $0.43$ \tabularnewline
    \hline
  \centering $1$ & $1$ & $1$ & $0.9$ & $0.9$ & $1$ & $1$\tabularnewline
    \hline
  \end{tabular}
\end{table}

\subsubsection*{\textbf{$2$-input $4$-output balanced  AND/OR gates}}
\begin{figure}[t]
\centering
  \includegraphics[width=\linewidth]{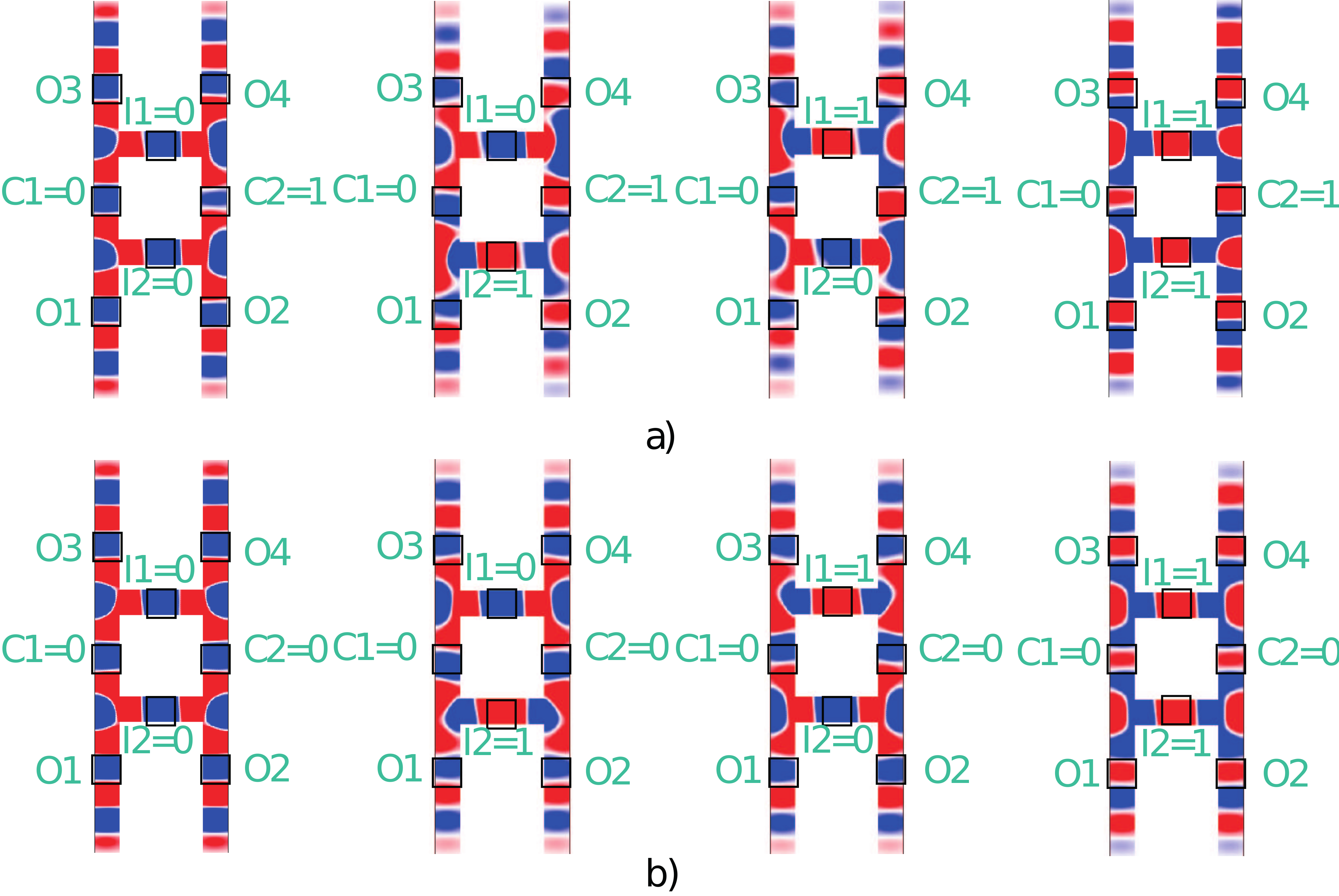}
  \caption{$2$-input $4$-output balanced a) AND/OR Gate b) AND/AND Gate}
  \label{fig:results4}
\end{figure}

The simulation results for $2$-input $4$-output balanced AND/OR gate for $4$ cases $I_1I_2$ $=$ {$00$ $01$ $10$ $11$} are presented in Figure \ref{fig:results4}a. The simulation results of the $2$-input AND/AND gates are presented in Figures \ref{fig:results4}b. By inspecting Figure \ref{fig:results4}a, the left arm provides the AND gate functionality in the two outputs $O_1$ and $O_3$.  On the other hand, the right arm provides the OR gate results in the two outputs $O_2$ and $O_4$. These are placed with $O_1$ and $O_3$ at distances $d_1$$=$$d_2$$=$$d_7$$=$$d_8$$=$$110$nm (n=1). The same line of thinking as the previous $2$-input cases can be followed to analyze the results. Taking $O_1$ and $O_3$ as an example, if the inputs are $I_1I_2$$=$$00$,  $I_1I_2$$=$$01$,  $I_1I_2$$=$$10$, then the output becomes $O_1$$=$$0$ and $O_3$$=$$0$. Also, $O_1$$=$$1$ and $O_3$$=$$1$ for the input combination $I_1I_2$$=$$11$. The OR gate result is obtained from $O_2$. In addition, NAND and NOR gates can be captured by placing the reading positions at $\lambda/2$, i.e., $d_1$$=$$d_2$=$d_7$$=$$d_8$=$55$nm (n=0). Therefore, the structure can provide AND, NAND, OR, and NOR gates. Likewise, Figure \ref{fig:results4}b can be analyzed. Also,  $2$-input (N)OR/(N)OR gates can be obtained in the same manner but with $C_1$$=$$C_2$$=$$1$. Thus, the structure can provide AND, NAND, OR, and NOR gate functionalities and each gate column isable to provide  AND (OR) in its direct and inverted format or in the same format with a fanout of $2$.

\subsubsection*{\textbf{$2$-input $4$-output balanced XOR/XNOR gates}}
Table \ref{table:6} presents the normalized MSAs at the outputs $O_1$, $O_2$, $O_3$, and $O_4$ for $C_1$$=$$C_2$$=$$0$ and $C_1$$=$$C_2$$=$$1$ and for different inputs combination $I_1I_2$ $=$ {$00$ $01$ $10$ $11$} for the balanced $4$-output structure. Note that the cases $C_1 C_2$ $=$ {$01$ $10$} results are exhibiting the same behaviour.
Table \ref{table:6} indicates that XOR and XNOR can be now implemented at all four outputs by making use of the same threshold value $0.38$ obtained by averaging the normalized $O_1$, $O_2$, $O_3$, and $O_4$ MSA for input combinations 01 and 11. To implement the XOR gate, the condition must be: if the normalized MSA is larger than $0.38$, then outputs equal to logic $0$ and logic $1$ otherwise. The XNOR gate can be captured by flipping the condition. Therefore, the structure can provide different combinations of XOR, XNOR and enable a fanout value up to $4$.

\begin{table}[t]
\caption{$2$-input $4$-output Balanced Gate Normalized Outputs MSAs}
\label{table:6}
\centering
  \begin{tabular}{|c|c|c|c|c|c|c|}
    \hline
   \multicolumn{3}{|c|}{Cases} &  $O_1/I_1$  & $O_2/I_1$  & $O_3/I_1$  & $O_4/I_1$ \tabularnewline \hline
    $C_1=C_2$ & $I_2$& $I_1$& & & & \tabularnewline
    \hline
     $0$ & $0$ & $0$ & $1$ & $1$ & $1$ & $1$\tabularnewline
    \hline
     $0$ & $0$ & $1$ & $0.33$ & $0.33$ & $0.33$ & $0.33$\tabularnewline
    \hline
     $0$ & $1$ & $0$ & $0.3$ & $0.3$ & $0.3$ & $0.3$ \tabularnewline
    \hline
     $0$ & $1$ & $1$ & $0.43$ & $0.43$ & $0.43$ & $0.43$ \tabularnewline
    \hline
    $1$ & $0$ & $0$ & $0.43$ & $0.43$ & $0.43$ & $0.43$ \tabularnewline
    \hline
     $1$ & $0$ & $1$ & $0.3$ & $0.3$ & $0.3$ & $0.3$ \tabularnewline
    \hline
     $1$ & $1$ & $0$ & $0.33$ & $0.33$ & $0.33$ & $0.33$\tabularnewline
    \hline
     $1$ & $1$ & $1$ & $1$ & $1$ & $1$ & $1$\tabularnewline
    \hline
  \end{tabular}
\end{table}

\subsection{Discussion}
\label{subsec:Discussion}
In the sequel, the proposed PLG is assessed and compared with the state-of-the-art SW and $16$ nm CMOS counterparts. In addition, fanout achievement, geometric scalability, balanced spin wave strength, and variability and thermal noise effects are discussed.

\subsubsection*{\textbf{Comparison}}
We evaluated the proposed $4$-output PLG structure in terms of energy and delay, and compare it with state-of-the-art SW \cite{Excitation_table_ref16} and $16$ nm CMOS \cite{16nmCMOS} functionally equivalent designs. We followed the  assumptions made in \cite{Excitation_table_ref16} to make a fair comparison: (i) SW excitation and detection cells are ME cells, which have an area of $48$~nm $\times$ $48$~nm, (ii) pulse signals are used to excite spin waves, (iii) No energy and delay are accounted for the output ME cell because the structures output are fed to the following SW gates, (v) $0.42$~ns ME cell switching delay, $C_{ME}=1$ fF, $V_{ME}=119$ mV, Energy=$I \times C_{ME} \times V_{ME}^2$ (where $I$ is the number of excitation cells), and SW $\lambda = 48$~nm, (vi) The SW propagation delay is negligible. Note that the made assumptions might not reflect the reality of the current spin wave based technology due to the early stage development of the technology, but their discussion is not part of this paper. 

Moreover, we assumed that AND, OR, XOR, and XNOR $16$ nm CMOS logic gates constitute CMOS PLG. Also, the energy and delay numbers were estimated based on the energy and delay numbers for the logic gates, which were taken from \cite{16nmCMOS}.  

Our evaluation results are presented in Table \ref{table:6}. As it is clear form the Table, compared to $16$ nm CMOS, the proposed gate is $11$x slower and consumes $16$x less energy. In addition, the design in \cite{Excitation_table_ref16} is performing slightly better in performance, but the Majority gate in \cite{Excitation_table_ref16} can provide maximally one output. Therefore, if more outputs are needed, the circuit must be replicated multiple times, thus needs more energy. For instance, when using the design in \cite{Excitation_table_ref16}, if the output is needed $4$ times the structure must be replicated $4$ times leading to an energy consumption of $173$ aJ. Our $4$-output structure consumes $57.6$ aJ,  therefore it needs $3$x less energy for the same computation without encoring  any delay overhead. 

\begin{table}[t]
\caption{Comparison with SW and CMOS}
\label{table:7}
\centering
  \begin{tabular}{|c|c|c|c|}
   \hline
     & PLG CMOS  &  MAJ gate SW & Proposed SW PLG \tabularnewline \hline
     Technology &  16 nm CMOS &  SW & SW \tabularnewline
    \hline
     Implemented function &  AND, OR, XOR & MAJ gate & (N)AND, (N)OR, X(N)OR \tabularnewline
    \hline
     Number of used cell & $26$ transistors &  $4$ ME cells & $8$ ME cells \tabularnewline
   \hline
     Max. No. output capability & \textgreater $4$ &  $1$ & $4$ \tabularnewline
    \hline
     Delay (ns) &  $0.047$ &  $0.42$ & $0.42$ \tabularnewline
    \hline
     Energy (aJ) &  $923$ &  $43.3$ & $57.6$ \tabularnewline
    \hline
  \end{tabular}
\end{table}

\subsubsection*{\textbf{Fanout Achievement}}
\begin{figure}[t]
\centering
  \includegraphics[width=\linewidth]{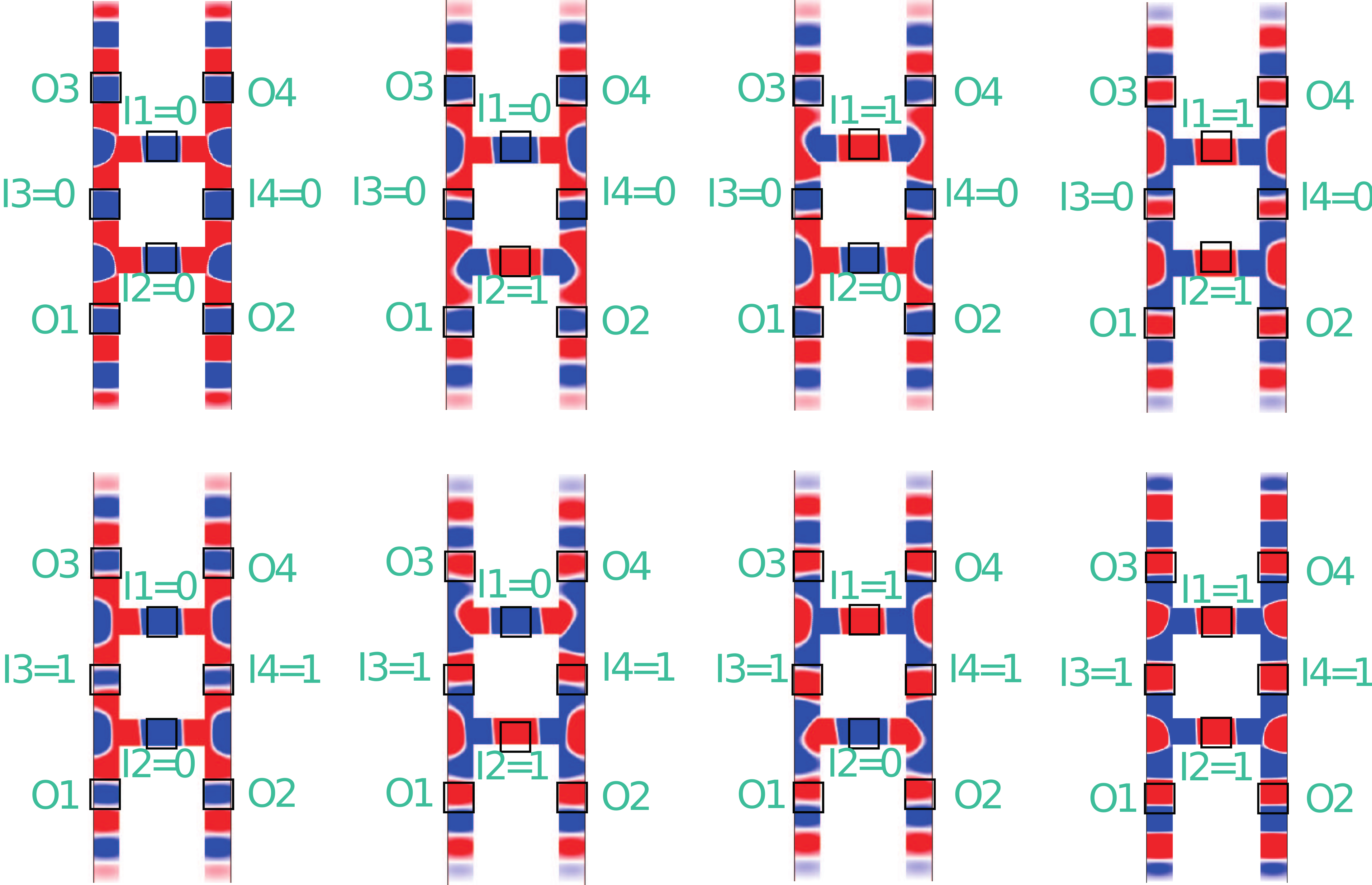}
  \caption{Fanout of $4$ $3$-input Majority Gate Simulation Results. SW phase is color encoded:  Red presents logic $1$ and blue logic $0$. }
  \label{fig:results5}
\end{figure}

If not only multi-output is desired, but also fanout capability, then all outputs must have the same energy level. The $4$-output structure presented in Figure \ref{fig:FO4}, can achieve $2$ times a fanout of $2$ but not a fanout of $4$ because the normalized $O_1$, $O_2$, $O_3$, and $O_4$ MSA are not the same in all cases as presented in Table \ref{table:5}. In contrast, the balanced $4$-output structure depicted in Figure \ref{fig:OEBFO4} has fanout of 4 capability because all outputs have the same MSAs as indicated in Table \ref{table:6}. Also, as an additional example, we used the proposed PLG to implement a fanout of $4$ $3$-input Majority gate. The simulation results for this implementation are presented in Figure \ref{fig:results5}. By inspecting the Figure, the outputs $O_1$, $O_2$, $O_3$, and $O_4$ are the same for all input cases. The same line of thinking as the previous cases can be followed to analyze the results. If inputs $I_1I_2I_3$$=$$000$,  $I_1I_2I_3$$=$$001$,  $I_1I_2I_3$$=$$010$, and $I_1I_2I_3$$=$$100$, then the outputs are $O_1$$=$$0$, $O_2$$=$$0$, $O_3$$=$$0$, and $O_4$$=$$0$. Also, $O_1$$=$$1$, $O_2$$=$$1$, $O_3$$=$$1$, and $O_4$$=$$1$ for the input combinations $I_1I_2I_3$$=$$011$,  $I_1I_2I_3$$=$$101$,  $I_1I_2I_3$$=$$110$, and $I_1I_2I_3$$=$$111$. Thus the Majority behaviour is delivered and as according to Table \ref{table:6} all outputs exhibit the same energy level a fanout of $4$ $3$-input Majority gate is achieved.

\subsubsection*{\textbf{Balance Spin Wave Strength}}
It was observed that both control and data inputs have unbalanced contribution to the output. The path for SWs originating from  data input contains a bent, which is associated with additional energy loss. The path of SWs originating from the control inputs is straight and thus less energy is lost. Therefore, data inputs have smaller contribution to the outputs. As a result, in order to balance the contribution of the inputs to the outputs, the data inputs must be excited at higher energy than the control inputs $C_1$ and $C_2$. In addition, it was noticed that all outputs are affected by all inputs such that $O_2$ is affected by control inputs $C_1$ and $O_1$ is also affected by $C_2$. This might create wrong results when different functions are desired at $O_1$ and $O_2$. Therefore, in order to have a working gate, it must be ensured that $C_2$ has less effect on output $O_1$ when compared to the combined effect of $C_1$, $I_1$, and $I_2$. Furthermore, it must be ensured that the contribution of $C_2$, $I_1$, and $I_2$ on output $O_2$ is larger than the contribution of $C_1$.

\subsubsection*{\textbf{Variability and Thermal Noise Effect}}
Our main target in this paper is to validate the proof of concept of the proposed structures, regardless of the variability and the thermal noise effect. However, in \cite{DC,DC9}, edge roughness and trapezoidal cross section of the waveguide were presented to test their effect on the gate functionality. It was demonstrated that the gate functions correctly under their presence and they only have a small effect \cite{DC,DC9}. Furthermore, the thermal noise effect was analized in \cite{DC} and concluded that it has a small effect and the gate correct functionality at different temperatures. Thus, we don't expect a noticeable effect of variability and thermal noise on the proposed structures. However, the investigation of such phenomena is subject of future work.

\section{Conclusions}
\label{sec:Conclusion}

In conclusion, a novel ladder shaped $2$-input $4$-output programmable logic gate structure was proposed. We introduced the gate structure and demonstrated that, by adjusting the gate output detection method, it can parallelly evaluate any $4$-element subset of the $2$-input Boolean function set \{AND, NAND, OR, NOR, XOR, and XNOR\}. Furthermore, we adjusted the structure such that all its $4$ outputs produce SWs with the same energy and demonstrated that it can evaluate Boolean function sets while providing fanout capabilities ranging from $1$ to $4$. We validated our approach by instantiating and simulating different gate configurations such as $4$-output AND/OR, $4$-output XOR/XNOR, output energy balanced $4$-output AND/OR, and output energy balanced $4$-output XOR/XNOR by means of Object Oriented Micromagnetic Framework (OOMMF) simulations. We evaluated the performance of our proposal in terms of delay and energy consumption and compared it against existing state-of-the-art SW and $16$~nm CMOS counterparts.  The results indicated that, for the same functionality, our approach provides $3\times$ and $16\times$ energy reduction, when compared with conventional SW and CMOS implementations, respectively.

\section*{Acknowledgement} 
This work has received funding from the European Union's Horizon 2020 research and innovation program within the FET-OPEN project CHIRON under grant agreement No. 801055. It has also been partially supported by imec's industrial affiliate program on beyond-CMOS logic. F.V. acknowledges financial support from Flanders Research Foundation (FWO) through grant No.~1S05719N. 

\bibliography{2-input_4-output_Programmable_Spin_Wave_Logic_Gate}

\end{document}